\documentclass[twocolumn,twoside,slac]{revtex4}
\usepackage{graphicx}
\usepackage{fancyhdr}
\begin{document}

\title{A Secure Infrastructure For System Console and Reset Access}

%

\author{Andras Horvath, Emanuele Leonardi, Markus Schulz}
\affiliation{CERN IT/ADC, Geneva, Switzerland}

\begin{abstract}
During the last years large farms have been built using commodity
hardware. This hardware lacks components for remote and automated
administration. Products that can be retrofitted to these systems are
either costly or inherently insecure. We present a system based on
serial ports and simple machine controlled relays. We report on
experience gained by setting up a 50-machine test environment as well as
current work in progress in the area.
\end{abstract}

\maketitle

\thispagestyle{fancy}


\section{Area of operation}

Our LHC Grid system consists of several thousand\footnote{Current estimates
show that our farm will grow to about 6000 machines by 2007.} `worker nodes',
namely Linux-based PCs, connected via Ethernet network and used for large
distributed calculations. These \emph{nodes} have different designated
administrators who reinstall them rather often according to current user needs.
It is mainly during installation or when system administrators experiment with
new kernel patches etc.\ that \emph{console access} is most important.

Moreover, these computers can not be considered reliable and experience
shows that sometimes PCs tend to `hang' without any reproducible
reason. In this case (or when, for example, the kernel encounters an
internal programming error) nodes may need to be restarted by some
hardware means, e.g.\ pushing their \emph{reset button}.

Up to now both these operations have been done `by hand', e.g.\ 
by having to go physically to our Computer Center 
to connect a monitor/keyboard to the nodes in question or to push their reset
button. In addition to being inconvenient and time-consuming,
this practice increases the possibility of a human mistake.

We have moved the machines' console from the VGA screen and keyboard to using a
serial line. Serial console systems are easy to implement, have been long since
supported by Linux as well as being a `tradition' in the computer industry.

The drawback of having a serial console instead of connecting a monitor
to the node is that most BIOSes are unable to operate on a serial
line, therefore at least the Linux boot loader has to start before
the serial line can be used as console.

On the other hand, a `normal' PC console provides an interface for human access
only. But why is machine control important?  For example, an automated
monitoring system could restart non-responding machines a number of times
(within a given time period) before raising an alarm, thus reducing the
necessary amount of human intervention in case of `spontaneous hangs'. An
automated installation system could watch over the console of the nodes being
installed and take appropriate action based on pattern-matching on console
messages.

Our system offers controlled remote access for administrators to the consoles
and reset buttons of the nodes they are in charge of.

\section{Implementation}

\subsection{Hardware architecture}

We wanted the smallest amount of and possibly the cheapest extra hardware
to be added. In principle, it would be possible to have two serial
ports in each computer and `daisy-chain' them using null-modem
serial cables so that one of the ports is the node's own console and
it acts as a \emph{console server} to one other node. Provided that
the console server is accessible via the network, console operations
can be performed on its client remotely.

Having as many console servers as worker nodes is not generally a
good idea though as, apart from possible scaling problems, taking
down e.g.\  ten daisy-chained nodes would mean that only the first one's
console remains accessible from the network (the console server of
which is outside these ten nodes). 

Therefore we utilised special 8-port PCI serial boards%
\footnote{Currently: ExSys EX-41098 (http://www.exsys.ch)}
 and have successfully installed up to three of them in the same PC,
which means that 24 ports can be served from a single
console server. These provide standard RS-232 DB9 serial lines similar
to the ones usually on PCs.

Reset button pressing is implemented using serial-line-controlled
programmable relay boxes. These have eight relay ports each and can be
cascaded into a chain of eight boxes, allowing up to 64 potential clients
controlled via a single \emph{reset server} serial port. The relays are
programmed to give a one-second impulse to the reset connector on the
machines' mainboard.

\subsection{Software and user interface}

Users connect to a single web/application server using any HTTPS and
X.509 capable browser. Authentication is done via X.509 user
certificates, issued by CERN's own CA, and authorisation information is
stored in the same database as other system data. This interface is where
\emph{reset requests} are submitted (e.g.\ `I want to reset this-and-that
node for a given reason') as well as where serial console interconnection
information is retrieved from the database. A user can then connect to
the console server responsible for given node at the push of a button
with a Java-based SSH client offered on the web page  or use his/her own
SSH program to do so.

On the console/reset servers themselves Linux file access control and
permissions are used to separate privileges. For each serial port a
local Linux user is created under the same name (e.g.\ user \texttt{ttyS5}
 for \texttt{/dev/ttyS5}) and its password is set so that the only way to
log in is via key-authenticated SSH.\@These users have their login shell
set to a very minimalistic terminal program and the permissions on the
TTY devices are set so as to make it accessible only to the
corresponding user.

RSA keys are used to let users in into the console server as the given
`tty-user'. The RSA public keys are stored in the database and
distributed to console servers whenever authorisation data changes.
This is work in progress and we consider using two keys for each
user a weak solution (see Section~\ref{newstuff} for the current status).

The same web interface is used for administrative tasks (adding new
users, modifying authorisation information etc).
A \emph{console client
detection} procedure can also be run from here which helps maintaining a
consistent interconnection database\footnote{Of course, reset button
interconnections can not be automatically detected.}. 

The system is scriptable through direct database queries (and SSH in
case of console access). We use Oracle 8i as database backend.

\subsection{Network security}

Our network is far too large to be considered trusted and, therefore,
encrypted and authenticated communication is in use.

The web interface can be reached via HTTPS.\@Both users and the server
hold a certificate issued by the CERN CA the validity of which is
checked.

Console services (from user's machine to console server) and internal
data communication between web server and console servers use SSH (protocol
version 2) with RSA authentication.

Secure connections to the database server are implemented using SSH
tunnels (we tried to use the built-in SSL capability of Oracle 8i on our
Linux boxes but failed).
The server's identity is ensured by checking
its host key.

\subsection{Costs (as of March 2003)}

\subsubsection*{Serial consoles:} 

8-port serial boards and cabling: 

CHF32.25 / node = \textbf{\$23.54 / node }

\emph{(compare: standard console MUX with SSH access: \$110-\$300
/ node)}

\subsubsection*{Remote reset:}

Hand-made cables, connectors, relay boxes:

CHF23.2 / node = \textbf{\$17 / node} 

\section{Work in progress\label{newstuff}}

The early deployment and test systems were considered a success as
`proof-of-concept' and we have learned much from the first version.
Now we have started the general integration of the system into CERN's already
working services.

We aim at the utilisation of a still-in-development system which
provides a generic solution, based on a central information store, to
the distribution (and update) of configuration files to our nodes.  Both
interconnection data and user authorisation information can be managed
this way thus eliminating the need for a separate database and a
separate access control method. 

That is, we distribute SSH key files for console access and small
text-based configuration files (about which host is on which TTY port,
needed for logging). The same distribution system will pass the results
of the automatic detection procedure, generated into a file, back to the
central information store.

Moreover, the system will not be as much `distributed' as we first aimed
at. Instead of using our PC farm nodes themselves to act as
console/reset servers for each other, we dedicate machines for
management purposes, one for each of our `rack' units (about 44 machines).
This solution is more costly as it needs an extra PC and special serial
hardware\footnote{Digi Acceleport Xem, for example (http://www.digi.com)} but these management
nodes will be used for purposes other than being console servers as
well. 

We have also dropped the web based management interface --- Linux command
line tools are going to be more easily scriptable. All configuration
etc.\ will be done through the generic data modification interface to the
central information store.

Our software was re-written in order to increase efficiency and simplify the
code (removing database access parts etc).
Certain parts though, especially most of the low-level serial port access and
client detection code were reused from the previous version.

Currently, major features include:

\begin{itemize}
\item Constant serial console logging (via syslog)
\item Automatic console client detection (on request)
\item Full integration into local environment, authentication/authorisation systems
\item High security --- controlled access to consoles and restarting
\item Command line interface for enhanced scriptability
\end{itemize}

In the near future, we plan to roll the system --- the above-mentioned second
version --- out on a larger scale.

\section{Contact information}

Please feel free to direct any questions or comments to \emph{Andras.Horvath@cern.ch}.


\begin{acknowledgments}
Guner Passage and Preslav Konstantinov wrote the initial version of the
software for controlling the relay boxes.
\end{acknowledgments}

\end{document}